\documentclass[pdflatex, sn-mathphys-num]{sn-jnl}

\usepackage{graphicx}
\usepackage{multirow}
\usepackage{amsmath, amssymb, amsfonts}
\usepackage{amsthm}
\usepackage{mathrsfs}
\usepackage[title]{appendix}
\usepackage{xcolor}
\usepackage{textcomp}
\usepackage{manyfoot}
\usepackage{booktabs}
\usepackage{algorithm}
\usepackage{algorithmicx}
\usepackage{algpseudocode}
\usepackage{listings}
\usepackage{comment}
\usepackage{float}

\begin{document}


\title[Article Title]{%
  Mechanistic Insights into the Oxygen Evolution Reaction on
  Nickel-Doped Barium Titanate via Machine Learning-Accelerated
  Simulations}

\author[1,2]{\fnm{Kajjana} \sur{Boonpalit} }\email{kajjana.b$\textunderscore$s21@vistac.ac.th}
\author*[1]{\fnm{Nongnuch} \sur{Artrith}}\email{n.artrith@uu.nl}
\affil[1]{\orgdiv{Materials Chemistry and Catalysis}, \orgname{Debye Institute for Nanomaterials Science, Utrecht University}, \orgaddress{ \postcode{3584 CG}, \state{Utrecht}, \country{the Netherlands}}}
\affil[2]{\orgdiv{School of Information Science and Technology}, \orgname{Vidyasirimedhi Institute of Science and Technology}, \orgaddress{\postcode{21210}, \state{Rayong}, \country{Thailand}}}


\abstract{%
Electrocatalytic water splitting, which produces hydrogen and oxygen through water electrolysis, is a promising method for generating renewable, carbon-free alternative fuels.
However, its widespread adoption is hindered by the high costs of Pt cathodes and IrO$_{x}$/RuO$_{x}$ anode catalysts.
In the search for cost-effective alternatives, barium titanate (BaTiO$_{3}$) has emerged as a compelling candidate.
This inexpensive, non-toxic perovskite oxide can be synthesized from earth-abundant precursors and has shown potential for catalyzing the oxygen evolution reaction (OER) in recent studies.
In this work, we explore the OER activity of pristine and Ni-doped BaTiO$_{3}$ at explicit water interfaces using metadynamics (MetaD) simulations.
To enable efficient and practical MetaD for OER, we developed a machine learning interatomic potential based on artificial neural networks (ANN), achieving large-scale and long-time simulations with near-DFT accuracy.
Our simulations reveal that Ni-doping enhances the catalytic activity of BaTiO$_{3}$ for OER, consistent with experimental observations, while providing mechanistic insights into this enhancement.}


\maketitle

\section{Introduction}\label{sec1}


Electrocatalytic water splitting with proton exchange membrane (PEM) electrolyzers, a promising approach for producing clean hydrogen fuel, is hindered by the high cost of precious metal catalysts~\cite{Henergy1, cost1, cost2}.
At room temperature, platinum is essential for the hydrogen evolution reaction (HER) at the cathode, while iridium and ruthenium oxides (IrO$_{x}$/RuO$_{x}$) are required for the oxygen evolution reaction (OER) at the anode~\cite{Henergy1, WaterSplit1}.
This dependence on scarce and expensive metals drives up production costs and limits the scalability and commercial viability of water electrolysis for widespread hydrogen production~\cite{cost1, cost2}.
To address these challenges, significant research efforts are focused on the development of more cost-effective catalysts for both the HER and OER, aiming to make this sustainable energy solution more accessible and affordable~\cite{trend1, trend2}.

For the OER, the search for novel catalyst to replace iridium and ruthenium oxides remains an active area of research.
The 3d transition metal oxides, such as nickel oxide (NiO) and cobalt oxide (Co$_{3}$O$_{4}$), have emerged as alternatives due to their abundance, lower cost, and favorable catalytic properties~\cite{alterOER}.
Perovskite oxides, with the general formula ABO$_{3}$, are particularly noteworthy.
These materials offer a flexible crystal structure that can be engineered by substituting different elements at the A and B sites, enabling fine-tuning of their electronic and catalytic properties~\cite{perovskite1, perovskite2}.
Perovskite oxides such as lanthanum strontium cobalt oxide (LSCO) \cite{LaSrCo3} and lanthanum nickel oxide (LNO) \cite{LaNiO3} have shown exceptional performance as OER catalysts.
These oxide materials can be further modified by doping with other elements to enhance their conductivity and catalytic activity~\cite{dopereview, WaterSplit1, NiFe@BTO, dopeinfluence}.

In this work, we aim to study the catalytic activity of barium titanate (BTO, BaTiO$_{3}$) perovskite oxide materials.
Since BTO is inexpensive and non-toxic~\cite{BTOcheap, BTOtoxic}, BTO-based catalysts for water electrolysis would be highly desirable.
Water electrolysis over BTO electrodes has been reported~\cite{OER-BTO, OER-BTO2, OER-BTO3}, and its OER activity can be enhanced by Ni-doping (Ni@BTO, Ni@BaTiO$_{3}$)~\cite{OER-BTO3}.
This experimental observation is consistent with density functional theory (DFT) calculations, which show a reduction in the theoretical overpotential of Ni@BTO, compared to BTO~\cite{NiFe@BTO}.
The density of states for Ni@BTO also suggests an enhancement in the electrical conductivity of the material, making it suitable for electrocatalysis~\cite{NiFe@BTO}.
However, there is a lack of studies to determine the activation energy for OER, particularly over BTO and Ni@BTO.

Most theoretical calculations of OER do not account for finite temperature effects and an explicit water environment~\cite{OER-MTD}.
To simulate models with explicit water molecules, empirical force fields lack the accuracy required and do not account for water dissociation.
DFT calculations face limitations related to system size and time scale~\cite{OER-MTD2}.
Machine learning potentials (MLP) can help overcome these limitations by enabling sufficiently large-scale dynamic simulations with DFT-level accuracy.

In this study, we conducted MLP-based molecular dynamics (MD) and metadynamics (MetaD) simulations to explore the reaction mechanism and free energy surface of the OER over BTO and Ni@BTO, including an explicit water environment.
We developed and adopted our MLP model to facilitate large-scale and long-time MD and MetaD simulations, as the OER might not be observable in shorter simulation times.

\section{Results}\label{sec2}

\subsection{MLP Training}\label{subsec2.1}

\begin{figure}[h!]
\centering
\includegraphics[width=1\textwidth]{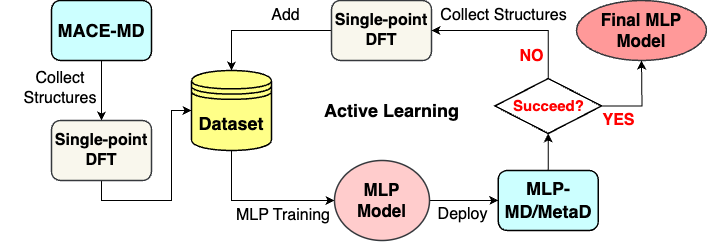}
\caption{Overview of MLP training pipeline, including dataset generation, data labeling, model training, model deployment, and active learning}\label{fig1}
\end{figure}

In this work, we aim to use the MLP to facilitate the study of the OER through MetaD simulations. Our MLP training pipeline is illustrated in Figure~\ref{fig1}. To train the MLP for this specific application, we built a custom dataset that covers the configurational space of OER over BTO and Ni@BTO slabs. The diverse range of structures in the training set is crucial for ensuring a broad applicability domain of the MLP. \cite{AD} Ab-initio molecular dynamics (AIMD) has been a common tool for dataset generation; however, this approach is a bottleneck and cannot provide configurations over long simulation times.
Herein, we used classical MD simulations combined with the MACE-mp-0 model \cite{MACE,mace-mp} to generate the configurational space. MACE-mp-0 is a universal MLP, trained to predict the energy and forces of a large materials database at the PBE \cite{PBE} level of theory. Its qualitative accuracy has been validated across various chemical systems, demonstrating satisfactory performance. \cite{mace-mp} This approach enabled us to generate configurations with DFT-level quality over longer simulation times, while significantly reducing the runtime. Without performing DFT calculations at every AIMD step, the energy and forces of selected generated structures can be recalculated using single-point DFT calculations with a preferred functional. This approach offers improved parallelization compared to the inherently sequential nature of AIMD.

Incomplete knowledge of the configurational space by the MLP model can lead to artifactual collapse of structures during simulations. The simulation diverged, exhibiting a nonphysical surge in temperature and potential energy, which caused the trajectory to deviate in an unrealistic manner. \cite{kong2023overcoming} To resolve this, we implemented an active learning (AL) scheme to gather unseen structures, enhancing the model's understanding of the configurational space. In AL, we run the MLP-MD and MLP-MetaD simulations to collect unseen configurations that occur at very long MD timescales and at the transition states of the OER, which were not covered by the MACE-generated initial dataset. Although the collapsed structures are not preferred configurations, their presence in the training dataset is valuable for improving the model's understanding of the complete configurational space. \cite{negative,negative2} We repeated the AL loops until we were able to successfully complete the MD and MetaD simulations with stability, without any structural collapse. Consequently, our final dataset consists of 16,162 structures, which their energy and forces are recalculated with RPBE+D3 functional. \cite{RPBE,D3} The details of dataset are provided in Appendix~\ref{secA1}.

\begin{figure}[h!]
\centering
\includegraphics[width=1\textwidth]{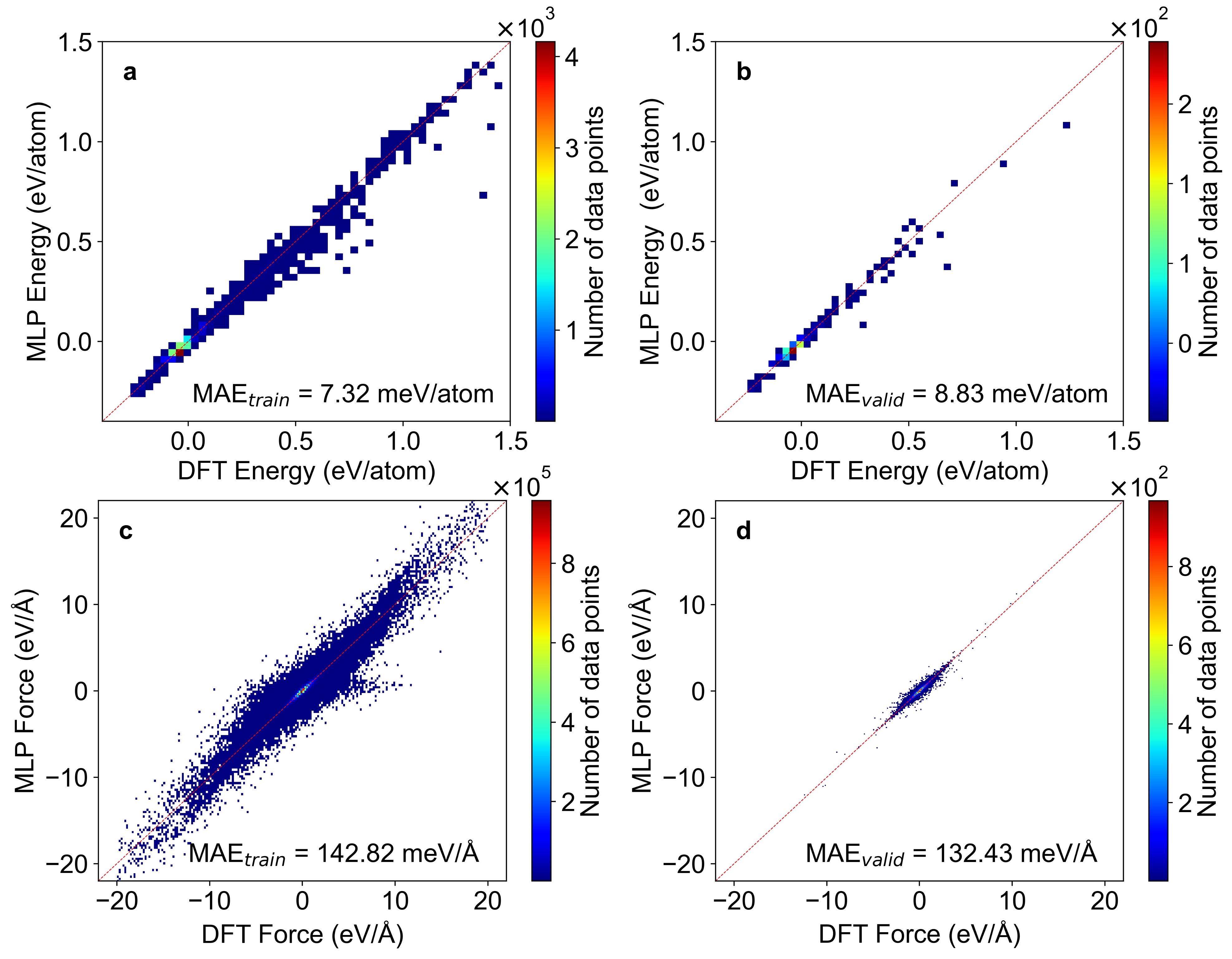}
\caption{The parity plots display the agreement between energies and forces derived from DFT calculations and MLP predictions, with the color representing the density of data points. (a-b) show the energy correlation between DFT and MLP for the training and validation datasets, respectively, while (c-d) illustrate the force correlation between DFT and MLP for the training and validation datasets.}\label{fig2}
\end{figure}

Tables \ref{tabB1} and \ref{tabB2} report the benchmarking results of the MLP models, including the prediction errors for energies and forces. Our best MLP model utilizes a 30-20-20-20-1 artificial neural network (ANN) architecture. It exhibits mean absolute errors (MAEs) of 7.32 meV/atom for energies and 142.82 meV/Å for forces in the training set. Meanwhile, the energies and forces MAEs of validation set are 8.83 meV/atom and 132.43 meV/Å, respectively. Figure~\ref{fig2} demonstrated the correlation between DFT ground truths and MLP predictions, showing a good agreement between MLP and DFT.

Additionally, we assessed the performance of trained MLP by evaluating its ability to replicate the structural properties observed in AIMD trajectories. The MLP can effectively capture the structural configuration. Figure~\ref{figB1} demonstrate the agreement in the radial distribution functions (RDFs) between AIMD and MLP-MD simulations, both conducted with the same simulation box: BTO(2$\times$2)/32H$_2$O and Ni@BTO(2$\times$2)/32H$_2$O, respectively.

After the MLP validations, the trained MLP is used for the MetaD study of the OER at the BTO/water and Ni@BTO/water interfaces. The BTO(4$\times$4)/128H$_{2}$O and Ni@BTO(4$\times$4)/128H$_{2}$O systems (Figure~\ref{figC3}) are equilibrated at 300 K for 500 ps using MLP-MD simulations. The density profiles of water (Figure \ref{figB2}a) demonstrate that the bulk water has a density of 1 g/mL. The resulting thermally equilibrated configurations were utilized as the initial states for the MLP-MetaD simulations.

\subsection{Metadynamics Study of OER}\label{subsec2.2}

The conventional OER mechanism, involving four-electron transfers, can be expressed as a stepwise reaction as follows:
\begin{equation}
H_{2}O_{(l)} + ^{*} \rightleftharpoons  OH^{*} + H^{+} + e^{-}
\end{equation}
\begin{equation}
OH^{*} \rightleftharpoons  O^{*} + H^{+} + e^{-}
\end{equation}
\begin{equation}
OH^{*} + H_{2}O_{(l)} \rightleftharpoons  OOH^{*} + H^{+} + e^{-}
\end{equation}
\begin{equation}
OOH^{*} \rightleftharpoons  O_{2}^{*} + H^{+} + e^{-}
\end{equation}
Nonetheless, previous DFT studies of the OER have predominantly concentrated on the thermodynamic aspects of the reaction, often overlooking critical kinetic information at the TS. Additionally, many of these studies neglect the influence of explicit solvent effects, which can significantly stabilize reaction intermediates. \cite{tripkovic2010oxygen,karlberg2006adsorption,briquet2017new}

\begin{figure}[h!]
\centering
\includegraphics[width=1\textwidth]{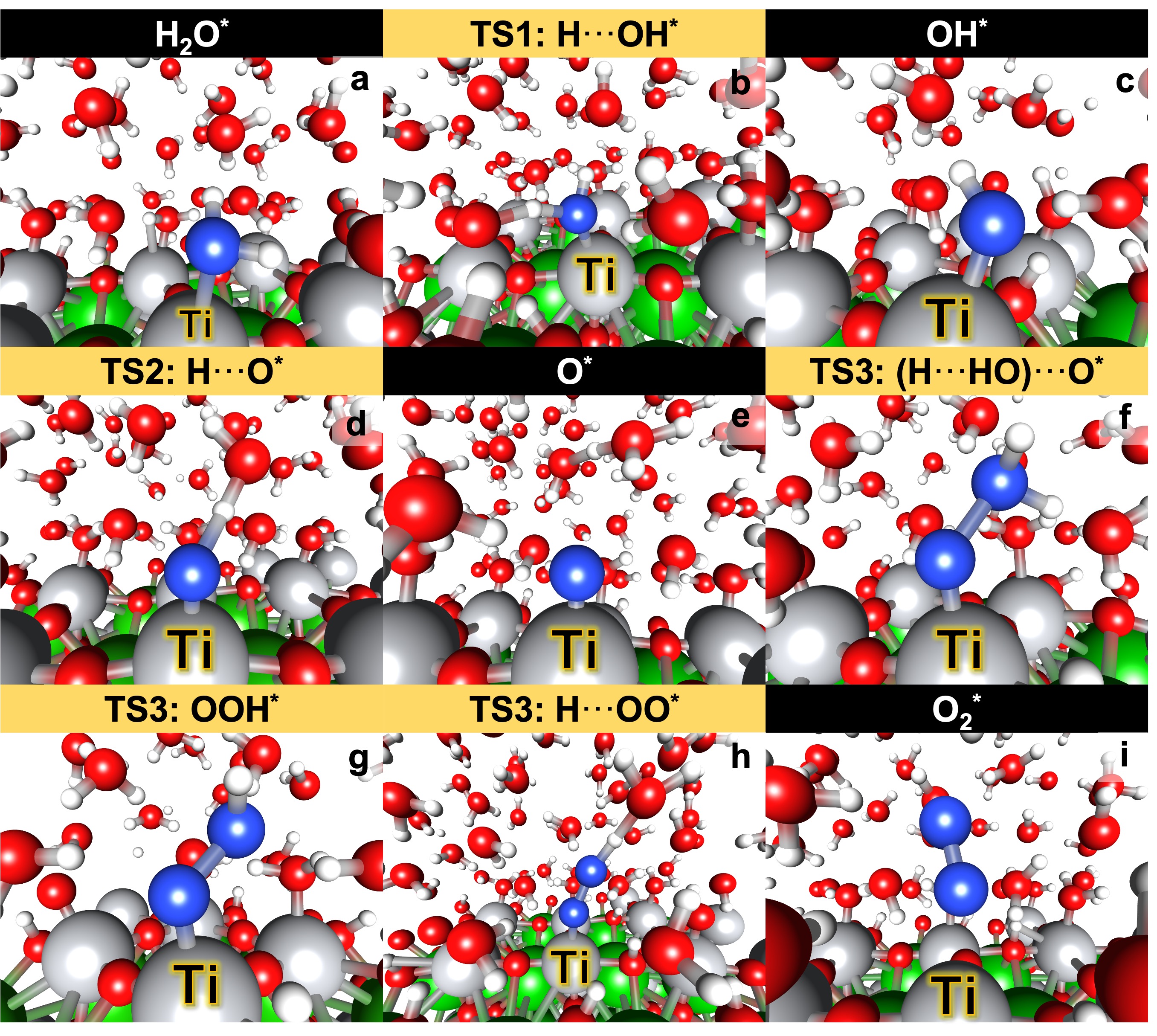}
\caption{The OER mechanism observed from the MLP-MetaD trajectory of BTO(4×4)/128H$_{2}$O system. The coloring scheme corresponds to the following atom types: titanium (grey), barium (green), hydrogen (white), oxygen (red), and reacting oxygen (blue). The Ti active site is annotated in the figure.}\label{fig3}
\end{figure}

\begin{figure}[!h]
\centering
\includegraphics[width=1\textwidth]{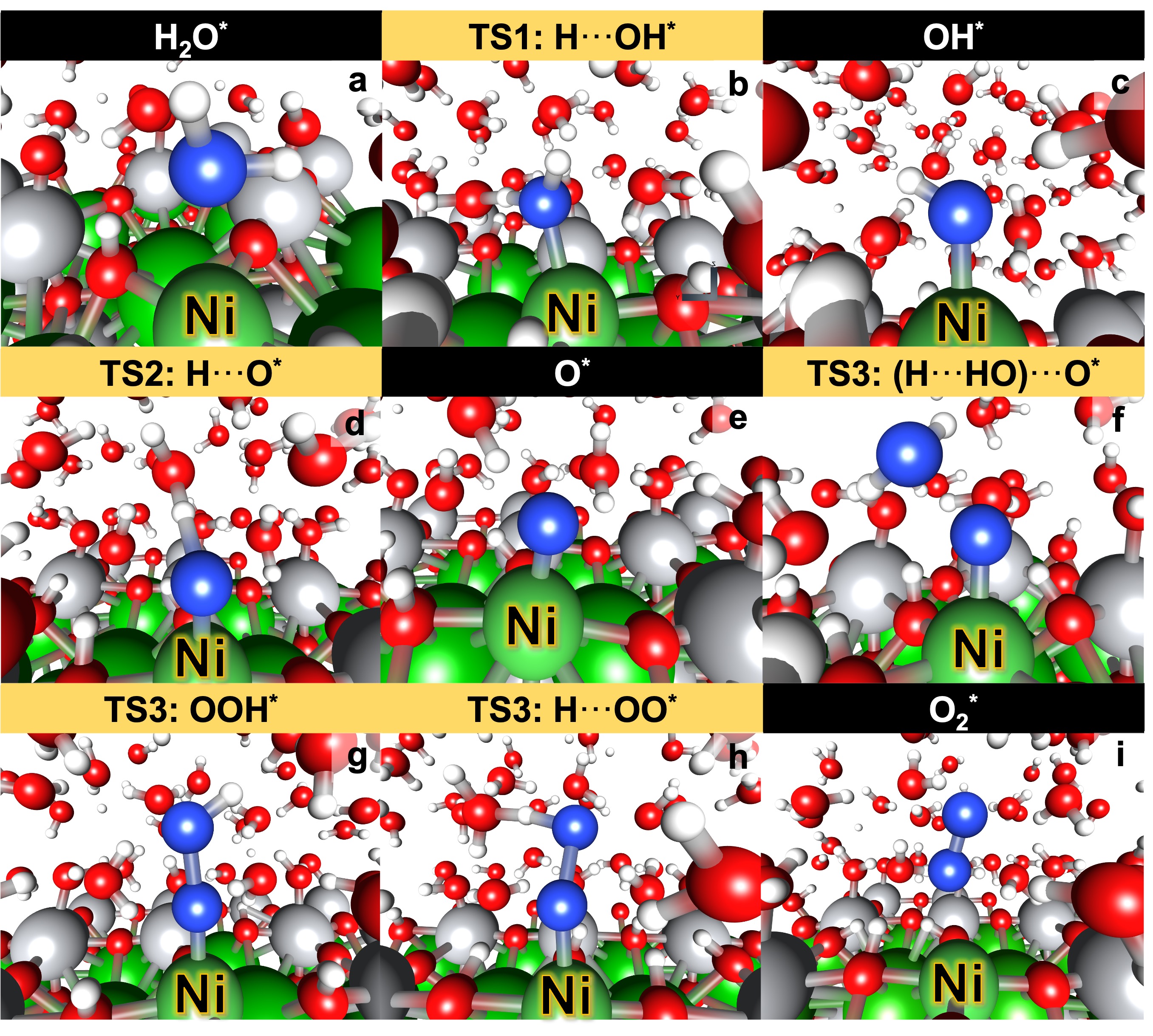}
\caption{The OER mechanism observed from the MLP-MetaD trajectory of Ni@BTO(4×4)/128H$_{2}$O system. The coloring scheme corresponds to the following atom types: titanium (grey), barium (green), nickel (light green), hydrogen (white), oxygen (red), and reacting oxygen (blue). The Ni active site is annotated in the figure.}\label{fig4}
\end{figure}

\begin{figure}[h!]
\centering
\includegraphics[width=1\textwidth]{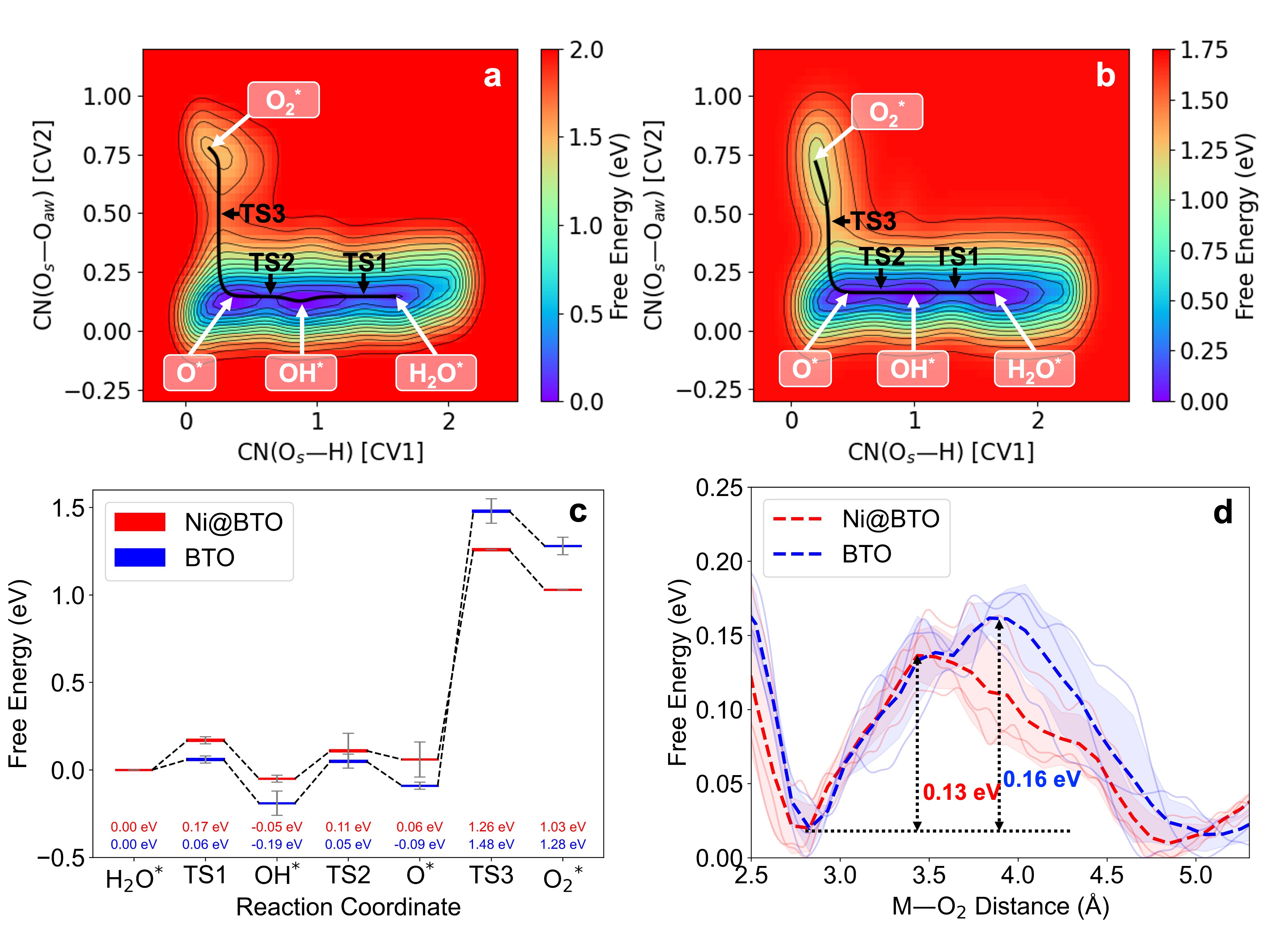}
\caption{The FES of OER over (a) BTO and (b) Ni@BTO derived from MetaD simulations with bias factor of 15, and Gaussian height of 0.05 eV. The black line represent the minimum energy path, corresponding to transitions from H$_{2}$O$^{*}$ to O$_{2}$$^{*}$. (c) The free energy profiles of OER over BTO (blue) and Ni@BTO (red). (d) The free energy profiles of O$_{2}$ desorption from surface. The free energy profiles were averaged over the FESs three MetaD runs, each starting from different initial configurations. For (d), the dashed line represents the average FES, while the shaded area indicates its standard deviation. The transparent solid lines correspond to the FES obtained from each individual run.}\label{fig5}
\end{figure}

To understand both the kinetics and thermodynamics of the OER over BTO and Ni@BTO, we analyzed the trajectories and free energy surface (FES) derived from MLP-MetaD simulations. For each system, independent MetaD runs were conducted with varying initial configurations and MetaD parameters to confirm the reproducibility of the FES and to ensure that the results are independent of the chosen initial configurations and MetaD parameters. The details of collective variables (CV) and MetaD parameters are provided in Section~\ref{subsec4.4}. In the following sections, we refer to the oxygen atom bound to the slabs at central Ti or doped-Ni atoms as O$_{s}$, the oxygen atom of all water molecules (except O$_{s}$) as O$_{aw}$, and the oxygen atom of the reacting water molecule as O$_{w}$. We summarized the reaction mechanism of OER over BTO and Ni@BTO in Figure~\ref{fig3}-\ref{fig4}, while their corresponding free energy surfaces and profiles are provided in Figure~\ref{fig5}. The movie of transition states (TS) in OER can be seen in Movies S1.

Step 1 (Figure~\ref{fig3}a--c and \ref{fig4}a--c): Given H$_{2}$O$^{*}$ is a reference point, the first step is the water dissociation onto the surface, generating OH$^{*}$ intermediates.
This water dissociation step proceeds via TS1, where the proton is abstracted by the solvent. For BTO, the free energy barrier ($\Delta$$G^{\ddagger}$) for this step ($\Delta$$G^{\ddagger}_{H_{2}O\to OH}$) is 0.06 eV. In comparison, for the Ni@BTO, the $\Delta$$G^{\ddagger}_{H_{2}O\to OH}$ is higher, at 0.17 eV. The reaction free energy ($\Delta$G) of this elementary step is exothermic on both surfaces, releasing energy upon the formation of OH$^{*}$. For BTO, the reaction is exothermic by 0.19 eV, whereas for Ni@BTO, it is slightly less exothermic, with an energy release of 0.05 eV. For the water dissociation step alone, the reaction is slightly more favorable on BTO compared to Ni@BTO.

Step 2 (Figure~\ref{fig3}c--e and \ref{fig4}c--e): This step involves the formation of an O$^{*}$ species by proton abstraction from the OH$^{*}$. The TS2 of BTO exhibits the $\Delta$$G^{\ddagger}_{OH \to O}$ of 0.24 eV., while the $\Delta$$G^{\ddagger}_{OH \to O}$ for Ni@BTO is lower with the value of 0.16 eV. The $\Delta$$G_{OH \to O}$ is endothermic by 0.10 eV. and 0.11 eV. for BTO and Ni@BTO, respectively.

Steps 3--4 (Figure~\ref{fig3}e--i and \ref{fig4}e--i): In our FES analysis, the coordination number CN(O$_{s}$-H) and CN(O$_{s}$-O$_{aw}$) are used as the CVs to separate the intermediate states on FES. Since we do not have prior of which water molecules will react with O$_{s}$, we are unable to bias and monitor the CN(O$_{w}$-H) from the beginning. This limitation makes it challenging to differentiate the OOH$^{*}$ and O$_2$$^{*}$ states on the FES. Step 3 (Figure~\ref{fig3}e--g and \ref{fig4}e--g) involves the formation of the OOH$^{*}$ species through oxo-oxo bond formation. This step occurs when CN(O$_{s}$-O$_{aw}$) is approximately 0.3. Following this, Step 4 (Figure~\ref{fig3}g--i and \ref{fig4}g--i) involves the abstraction of a proton from OOH$^{*}$ to form O$_{2}$$^{*}$. The proton abstraction occurs at a higher CN value, typically more than 0.5. Since this OOH$^{*}$ to O$_{2}$$^{*}$ transition is positioned on the downhill side of the TS3 barrier and $\Delta$$G^{\ddagger}_{OOH\to O_{2} }$ is barrierless, the local minima representing the OOH$^{*}$ state becomes indistinct.

Despite the indistinguishable OOH$^{*}$ state, the $\Delta$$G^{\ddagger}$ for proton abstraction is expected to be significantly lower than $\Delta$$G^{\ddagger}$ for oxo-oxo bond formation. The OOH$^{*}$ is an unstable transitory intermediates in this reaction coordinate, which quickly undergoes proton abstraction to form O$_2$$^{*}$. Moreover, the TS of these steps do not interfere on the FES. There is no any potential impact on its interpretation of oxo-oxo bond formation energy barrier. As a result, TS3 accounts for both Step 3 and Step 4 of OER. The calculated free energy barriers for this combined transition, denoted as $\Delta$$G^{\ddagger}_{O \to O_{2}}$, are 1.57 eV for BTO and 1.20 eV for Ni@BTO. The $\Delta$$G_{O \to O_{2}}$ is endothermic, with values of 1.37 eV for BTO and 0.97 eV for Ni@BTO.

\subsection{Metadynamics Study of O$_{2}$ Desorption}\label{subsec2.3}

O$_{2}$ desorption (O$_{2}$$^{*}$ $\rightleftharpoons$ O$_{2 (g)}$ + $^{*}$) can play a crucial role in the OER as it significantly influences the overall reaction kinetics and efficiency. \cite{o2desorpim,O2desorpIrO2} If O$_{2}$ desorption is sluggish, it can lead to the accumulation of products on the surface, blocking active sites and thus hindering the reaction rate. According to the previous work on IrO$_{2}$ by Binninger et. al. \cite{iro2desorp}, O$_{2}$ desorption step would be rate-determining step (RDS) for the conventionally OER mechanism due to high O$_{2}$ desorption barrier. Gauthier et al. \cite{O2desorp} studied the OER kinetics on rutile-phase surfaces of IrO$_{2}$, RuO$_{2}$, RhO$_{2}$, and PtO$_{2}$, reporting that the estimated O$_{2}$ desorption barrier ranges from 0.28 eV (RuO$_{2}$) to 0.98 eV (IrO$_{2}$).

In DFT studies, the deprotonation of OOH$^{*}$ and O$_{2}$ desorption are often combined into a single step. However, this can lead to misleading conclusions about the reaction energetics. For instance, in the study of  OER on IrO$_{2}$ by Ping et al. \cite{misleadingOER}, the OOH$^{*}$ deprotonation step is exergonic with $\Delta$G of --0.27 eV at a potential of 1.23 V, whereas the O$_{2}$ desorption step is strongly endergonic with $\Delta$G of 0.53 eV. When these two steps are lumped together, the combined $\Delta$G is only 0.26 eV, making the overall step appear much more feasible than it actually is. This simplification overlooks the significant energetic barrier associated with O$_2$ desorption. \cite{o2desorpim} In this section, we performed another MLP-MetaD simulations to study O$_{2}$ desorption from BTO and Ni@BTO surfaces.

Figure~\ref{fig5}d illustrates the O$_{2}$ desorption barriers averaged from multiple MetaD simulations for both BTO and Ni@BTO surfaces. The results indicate that O$_{2}$ binds weakly to both surfaces, with Ni@BTO displaying a slightly lower desorption barrier. For Ni@BTO, the desorption barrier ($\Delta$$G^{\ddagger}_{des}$) is approximately 0.13 eV, while the $\Delta$$G^{\ddagger}_{des}$ for BTO is slightly higher, at around 0.16 eV. The $\Delta$$G_{des}$ on both BTO and Ni@BTO surfaces is nearly neutral. Thus, O$_{2}$ desorption is not the RDS.

Accordingly, the oxo-oxo bond formation (TS3) is considered the RDS of the OER. The  $\Delta$$G^{\ddagger}_{RDS}$ are 1.57 eV for BTO and 1.20 eV for Ni@BTO, which correspond to the theoretical overpotential of 0.34 and -0.03 V, respectively. When comparing the experimental overpotentials from BTO hollow porous spheres, the addition of Ni to the BTO reduces the overpotential from 0.80 V to 0.46 V. \cite{OER-BTO3} A reduction of 0.34 V shows a close agreement between theory and experiment.

These results suggest that Ni-doping lowers both the activation energy and the endothermicity of the process, facilitating the formation of O$_{2}^{*}$ on the Ni@BTO surface. The findings from all FES are consistent and confirm that the OER over Ni@BTO is more kinetically and thermodynamically favorable than that over BTO, which concurs with the experimental results. \cite{OER-BTO3} Although both static and thermodynamically pure DFT calculations \cite{NiFe@BTO} and kinetically involved MLP-MetaD simulations reach the same conclusion that Ni@BTO is a more active OER catalyst, their underlying free energy profiles differ. Nonetheless, the dynamic and explicit water effects captured in MetaD could provide deeper insights into the catalytic processes, as they are based on simulations of more realistic models.

\section{Conclusion}\label{sec3}

In this study, we systematically investigated the OER and O$_{2}$ desorption processes on BaTiO$_{3}$ and Ni-doped BaTiO$_{3}$ surfaces using a combination of machine learning potentials and metadynamics simulations.
Leveraging the foundational model MACE-mp-0 as a data generator, we developed a machine learning potential specifically designed for this complex five-element system.
This enabled efficient and accurate simulations of the entire OER pathway, incorporating explicit solvent effects within a single metadynamics simulation.
Our results reveal that Ni-doping significantly enhances catalytic activity by reducing the free energy barrier for oxo-oxo bond formation compared to pristine BaTiO$_{3}$.
Nonetheless, it is important to note that this study does not include the lattice oxygen-mediated mechanism (LOM) of OER.
Future investigations incorporating LOM could offer further insights into the catalytic performance of these materials.
The development of database and machine learning potential for BaTiO$_{3}$ and Ni-doped BaTiO$_3$ presented here provide a strong basis for future studies of LOM and other complex catalytic pathways.
The methodology presented in this work can be extended to simulate complex chemical reactions occurring at electrode–electrolyte interfaces with an explicit solvent environment.

\section{Methods}\label{sec4}

\subsection{Density Functional Theory (DFT)}\label{subsec4.1}

The spin-polarized DFT calculations were performed using the Vienna ab initio simulation package (VASP) \cite{vasp1,vasp2,vasp3} with the projector augmented wave (PAW) pseudopotentials \cite{PAW} and the revised Perdew–Burke–Ernzerhof exchange–correlation functional \cite{RPBE} combined with Grimme empirical correction \cite{D3} (RPBE+D3). The Brillouin zone was sampled at $\Gamma$-point only. The plane-wave cut-off energy was 520 eV. The energies and forces convergence limit were 10$^{–5}$ eV and 0.02 eV/Å, respectively. Gaussian smearing was used with a smearing width of 0.05 eV.

The five-layer BaTiO$_{3}$(001) and Ni@BaTiO$_{3}$(001) slabs was built with both 2×2×1 and 4×4×1 supercells. (52 and 208 atoms, respectively) The initial structure models were obtained from previous work by Artrith and co-workers. \cite{NiFe@BTO} We included explicit solvent, 32 and 128 H$_{2}$O molecules, above 2×2×1 and 4×4×1 slabs to reproduce the water density of 1 g/cm$^{3}$. The total number of atom in the simulation boxes are 148 and 592 atoms, respectively. Both models was included in dataset for machine learning potential training, while only 4×4×1 model was used for metadynamics study. The bottom three layers were kept fixed throughout all simulations.

\subsection{Machine Learning Potential (MLP)}\label{subsec4.2}

The MLP models were trained using the open-source Atomic Energy Network (ænet-PyTorch) package. \cite{aenet-pytorch} The dataset consists of 16,162 configurations. The input of MLP model was Artrith-Urban-Ceder (AUC) inter-atomic descriptors. \cite{AUCdescriptor} Further details of dataset generation and interatomic descriptors are provided in Appendix~\ref{secA1}. The train-validation set ratio is 95:5. The MLP employed an ANN architecture, consisting of the multi-layer perceptron with a hyperbolic tangent (tanh) activation function. We used the AdamW optimizer with a learning rate of 5 $\times$ 10$^{–3}$ and a batch size of 128. The model was trained for 1,000 epochs, and the epoch with the lowest validation loss was selected as the final model. We define the loss function as the weighted-sum of energy (E) and force (F) error.

\begin{equation}
L = (1-\alpha)L_{E} + \alpha L_{f},
\end{equation}
where $\alpha$ is a free weight parameter. (0.5 in this study) $L_{E}$ and $L_{F}$ are defined as
\begin{equation}
L_{E} = \sqrt{\frac{1}{N}\sum_{i=1}^{N_{struc}}(E_{i}^{ANN}-E_{i}^{DFT})^{2}}
\end{equation}
and
\begin{equation}
L_{F} = \sqrt{\frac{1}{\sum_{i=1}^{N_{struc}}3N_{atom,i}}\sum_{i=1}^{N_{struc}}\sum_{j=1}^{N_{atom,i}}(F_{i,j}^{ANN}-F_{i,j}^{DFT})^{2}} ,
\end{equation}
where $N_{struc}$ is the number of structures in the dataset and $N_{atom,i}$ is the amount of atoms in the $i^{th}$ structure. The forces acting on each atom can be computed by taking the derivative of the total energy with respect to the coordinates of the atoms.

\subsection{Molecular Dynamics (MD) Simulations}\label{subsec4.3}
\subsubsection{Initial Dataset Generation}\label{subsubsec4.3.1}
During initial dataset generation, multiple independent MD simulations were conducted using the MD engine and Langevin dynamics implemented in the Atomic Simulation Environment (ASE) \cite{ase} Python library. We utilized MACE-mp-0, \cite{mace-mp} a pre-trained foundation model  parameterized for 89 chemical elements, and MACEcalculator in ASE to execute the MD simulations. The version of MACE-mp-0 model is 2024-01-07-mace-128-L2.model. The MD simulations were propagated for up to 50 ps, targeting temperatures of 300 K, 500 K, and 700 K, with a timestep of 0.5 fs and varying initial configurations.

\subsubsection{Production MD Simulations}\label{subsubsec4.3.2}
We utilized our MLP models to accelerate the production MD simulations by ænet-LAMMPS \cite{aemet-lammps} interface. The MD simulations were conducted under the canonical ensemble (NVT) using a Nosé–Hoover thermostat \cite{nose} at a temperature of 300 K for 500 ps, with a timestep of 0.25 fs to equilibrate the systems.

\subsection{Metadynamics (MetaD) Simulations}\label{subsec4.4}

We performed well-tempered MetaD \cite{MetaD} simulations using ænet-LAMMPS-PLUMED2 \cite{aemet-lammps,plumed2} interface under the canonical ensemble (NVT) using a Nosé–Hoover thermostat \cite{nose} at a temperature of 300 K.

To study OER, the coordination number (CN) between two groups of atom was chosen as the CV, including CN(O$_{s}$-H) [CV1] and CN(O$_{s}$-O$_{aw}$) [CV2]. O$_{s}$, O$_{aw}$, and H denoted an adsorbed oxygen atom on centered metal atom, all other oxygen atom from surrounding H$_{2}$O molecules, and all hydrogen atoms from surrounding H$_{2}$O molecules, respectively. CN is defined as
\begin{equation}
CN = \sum_{ij}^{}\frac{(1-\frac{r_{ij}}{r_{0}})^m}{(1-\frac{r_{ij}}{r_{0}})^n},
\end{equation}
where r$_{ij}$ is the distance between atoms i and j. m, and n were fixed to 6 and 12, respectively. The r$_{0}$ was set to 1.2 Å and 1.5 Å for CV1 and CV2 to distinguish the formation of O$_{2}$ molecule during MetaD. The MetaD were conducted for 500 ps with timestep of 0.25 fs, using a Gaussian width of 0.1 and a deposition rate of every 62.5 fs (250 steps). The Gaussian height and bias factors applied in the simulations were varied, as detailed in Table ~\ref{tabC1}. The free energy surfaces (FES) are calculated only until the O$_{2}$ formation event occurs. This approach excludes the contribution of aqueous O$_{2}$ configurations that appear at longer simulation times, which could otherwise introduce irrelevant features into the FES.

To study oxygen desorption, we selected a distance between the center of mass O$_{2}$ molecule and centered-metal atom (d(Ti-O$_{2}$) or d(Ni-O$_{2}$)) as a CV. The MetaD simulations were conducted for 300 ps with a timestep of 0.25 fs. The MetaD parameters used include a Gaussian height of 0.01 eV, a Gaussian width of 0.05, and a deposition rate of every 6.25 fs (50 steps). The bias factors applied were 1 for BTO and 1.1 for Ni@BTO. This range of parameters was chosen because the energy barrier for oxygen desorption is relatively low compared to the barriers associated with the OER. The initial configuration was sampled from the OER MetaD trajectory.

\backmatter

\bmhead{Supplementary information}

The online version contains supplementary material available at

\bmhead{Acknowledgements}

This work was funded by a start-up grant (Dutch Sector Plan) from Utrecht University awarded to N.A.
K.B.\ was supported by a VISTEC-NSTDA collaborative research and education scholarship, and Srimedhi scholarship.
The authors acknowledge the ænetone HPC, the Dutch National e-Infrastructure, and the SURF Cooperative for providing the computational resources.
The authors gratefully acknowledge discussions with Sarana Nutanong and Supawadee Namuangruk.


\section*{Conflict of interest}

The authors declare no conflicts of interest.

\section*{Data availability}

The database from the Density Functional
Theory (DFT) calculations can be obtained from the GitHub repository at https://github.com/atomisticnet/XXXXX

\pagebreak
\begin{appendices}

\section{Dataset Construction}\label{secA1}

The dataset construction process involved multiple stages. Initially, the dataset was generated using MACE-MD (Section~\ref{subsubsec4.3.1}). Following this, the dataset underwent refinement through active learning (AL), first with MLP-MD (Section~\ref{subsubsec4.3.2}) and subsequently with MLP-MetaD (Section~\ref{subsec4.4}). Each stage contributed to the progressive enhancement and augmentation of the dataset, resulting in a well-curated set of structures. The structures were uniformly sampled throughout the MD/MetaD trajectory. In addition, the transition state configurations were manually selected from the MetaD trajectory, ensuring a focused collection of critical configurations representing key transition states.

The details and the number of structures included in the dataset are provided in Table~\ref{tabA1}. The total number of energy data points is 16,162 structures. The force data points were filtered using a maximum force cutoff of 20.0 eV/Å, yielding in 15,099 structures. To generate the molecular representations for MLP training, all structures were converted to Artrith-Urban-Ceder (AUC) inter-atomic descriptors \cite{AUCdescriptor}, representing the local atomic environment and chemical species. The AUC descriptors are constructed by expanding radial and angular distribution functions (RDF and ADF) in a basis set of Chebyshev polynomials. The radial and angular cutoff radius were set to 5.0 Å for RDF and 4.5 Å for ADF, while the order of the Chebyshev polynomials was set to 8 and 5, respectively.

\begin{table}[h!]
\caption{The details and number of structures in the dataset obtained during each stages of dataset construction. Some hydrogen atoms near the slabs were randomly removed, and O$_{2}$ molecules were introduced to generate varied initial configurations and working conditions.}\label{tabA1}%
\begin{tabular}{@{}lccc@{}}
\toprule
Systems & Initial (MACE)  & AL-MD  &  AL-MetaD   \\
\midrule
BTO(2×2)/32H$_{2}$O         &  3,597 & 1,170 &   \\
BTO(2×2)/32H$_{2}$O+O$_2$    &  198  &  &   \\
BTO(2×2)/30H$_{2}$O+2OH    &  869  &  &   \\
BTO(2×2)/30H$_{2}$O+2OH+O$_2$&  173  &  &   \\
BTO(2×2)/28H$_{2}$O+4OH    &    & 88 &   \\
BTO(2×2)/64H$_{2}$O       &  & 486 &  \\
\midrule
Ni@BTO(2×2)/32H$_{2}$O       & 3,651 & 190 &  \\
Ni@BTO(2×2)/32H$_{2}$O+O$_2$  & 200  & 565  &  \\
Ni@BTO(2×2)/30H$_{2}$O+2OH  & 903  & 184 &  \\
Ni@BTO(2×2)/30H$_{2}$O+2OH+O$_2$  & 188  &  &  \\
Ni@BTO(2×2)/28H$_{2}$O+4OH  &      & 373 &  \\
Ni@BTO(2×2)/64H$_{2}$O       &  & 494 &  \\

\midrule
BTO(4×4)/128H$_{2}$O       &    &  232  & 947  \\
BTO(4×4)/125H$_{2}$O+4OH       &    &  184  &   \\
\midrule
Ni@BTO(4×4)/128H$_{2}$O    &    & 58 & 1172 \\
Ni@BTO(4×4)/125H$_{2}$O+4OH    &    & 181   &  \\
\midrule
4Ni@BTO(4×4)/128H$_{2}$O   &    & 59 &  \\

\midrule
Total &  9,779  & 4,264 & 2,119  \\
\botrule
\end{tabular}
\end{table}

\pagebreak
\section{Model Benchmarking}\label{secB1}

All ANN based MLP models were trained using the open-source Atomic Energy Network (ænet-PyTorch) package. \cite{aenet-pytorch} To select the network architectures we benchmark the MLP predictive performance with our final dataset as shown in Table~\ref{tabB1} and \ref{tabB2}. We are hesitant to use the MACE architecture due to its significantly higher computational time requirements as as shown in Table~\ref{tabB3}.


\begin{table}[h!]
\caption{Mean absolute errors (MAEs) and root mean squared errors (RMSEs) in meV/atom of the neural network energies for the training set and the validation set with different network architectures and features. The notation of network architectures represents the number of multi-layer perceptron with nodes per layer. The train-validation set ratio is 95:5. The fit used in the present work is shown in bold.}\label{tabB1}%
\begin{tabular}{@{}lccccc@{}}
\toprule
Architecture & Radius/Angular Cut-off (Å) & MAE$_{train}$  & MAE$_{valid}$ & RMSE$_{train}$  & RMSE$_{valid}$  \\
\midrule
36-20-20-1     &5.0/5.0 &  9.72  & 10.81  & 21.29  &  26.20\\
30-20-20-1     & 5.0/4.5 &   7.36  & 8.47  &17.16 & 22.52 \\
36-20-20-20-1    & 5.0/5.0 & 11.80 & 13.20 & 21.52 & 27.20  \\
\textbf{30-20-20-20-1}    & \textbf{5.0/4.5} &   \textbf{7.32} &  \textbf{8.83} & \textbf{17.54} & \textbf{23.60} \\
\botrule
\end{tabular}
\end{table}

\begin{table}[h!]
\caption{Mean absolute errors (MAEs) and root mean squared errors (RMSEs) in meV/Å of the neural network forces for the training set and the validation set with different network architectures and features. The notation of network architectures represents the number of multi-layer perceptron with nodes per layer. The train-validation set ratio is 95:5. The fit used in the present work is shown in bold.}\label{tabB2}%
\begin{tabular}{@{}lccccc@{}}
\toprule
Architecture & Radius/Angular Cut-off (Å) & MAE$_{train}$  & MAE$_{valid}$ & RMSE$_{train}$  & RMSE$_{valid}$  \\
\midrule
36-20-20-1     &5.0/5.0 &   148.94 &  146.71 & 241.46 & 238.25 \\
30-20-20-1     & 5.0/4.5 & 145.02   & 142.73  & 235.52 & 228.93 \\
36-20-20-20-1    & 5.0/5.0 & 147.74 &  145.39 & 241.58 & 235.26 \\
\textbf{30-20-20-20-1}    & \textbf{5.0/4.5} &   \textbf{142.82} &  \textbf{132.43} & \textbf{233.06} & \textbf{207.96} \\
\botrule
\end{tabular}
\end{table}

\begin{table}[h!]
\caption{Evaluation of required computational time for MD steps of BTO(4$\times$4)/128H$_{2}$O simulation box (timesteps/s), and model training (min/epoch) using different model architectures.}\label{tabB3}%
\begin{tabular}{@{}lcccc@{}}
\toprule
Architecture & Hardware & Resource & timesteps/s & min/epoch \\
\midrule
ANN (30-20-20-20-1) & Intel Xeon Gold 6338 &64 Core CPU  &  85.64 &  3.5 \\
MACE (small)    & NVIDIA A40 & 1 GPU  &  1.63  & 50 \\
MACE (large)    & NVIDIA A40 & 1 GPU  &  0.35  & Memory Leak   \\
\botrule
\end{tabular}
\end{table}

\begin{figure}[!h]
\centering
\includegraphics[width=1\textwidth]{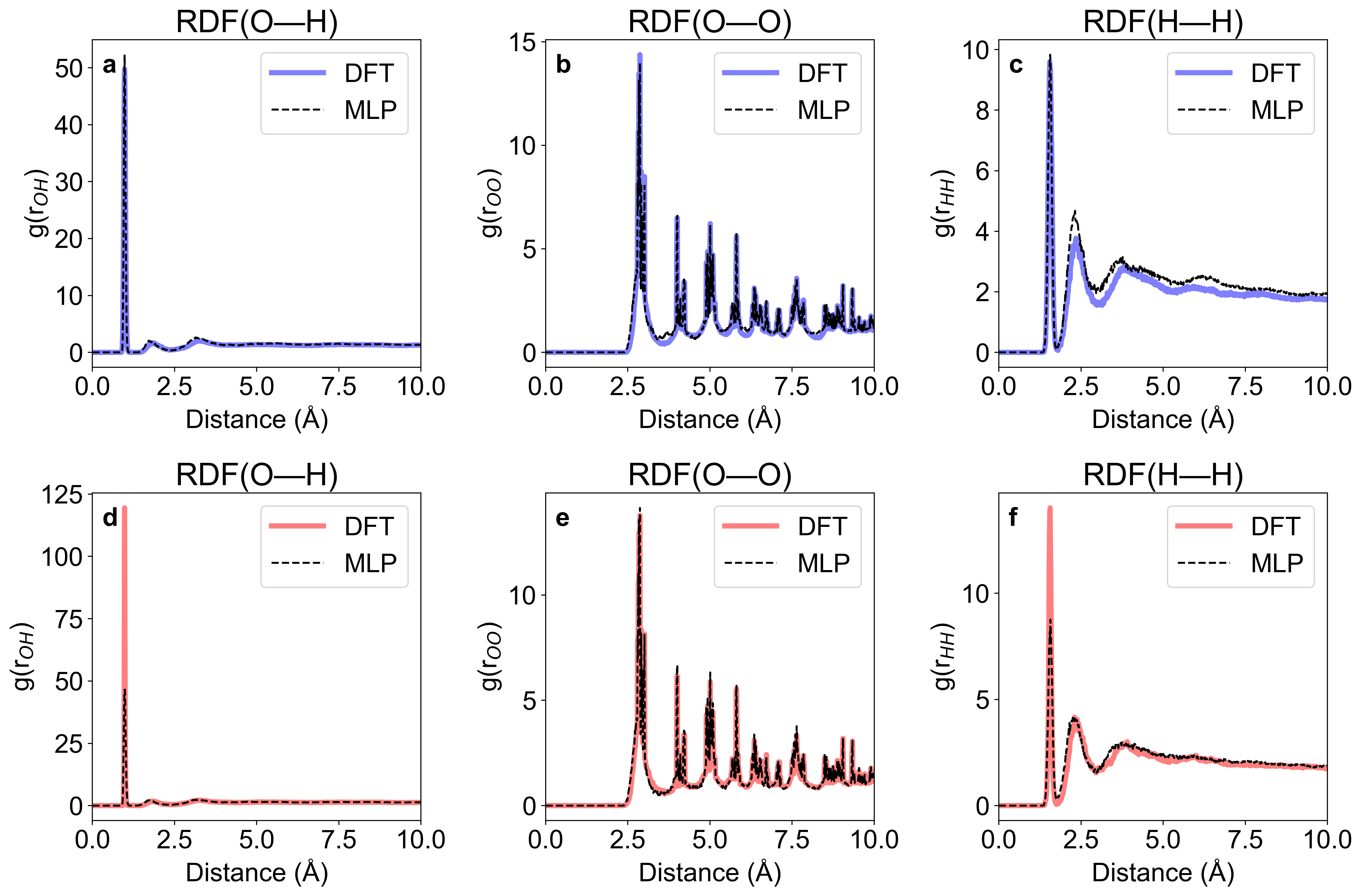}
\caption{Comparison between RDFs obtained from AIMD simulations and MLP-MD simulations of the (a-c) BTO(2x2)/32H$_{2}$O system and (d-f) Ni@BTO(2x2)/32H$_{2}$O systems. (a,d) RDF(O—H), (b,e) RDF(O—O), (c,f) RDF(H—H) The systems are equilibrated at 300 K for 15 ps, while the RDFs are averaged from the last 1 ps. We note that the simulations may not achieve equilibrium due to the limited affordable duration of the AIMD simulations.}\label{figB1}
\end{figure}

\begin{figure}[!h]
\centering
\includegraphics[width=1\textwidth]{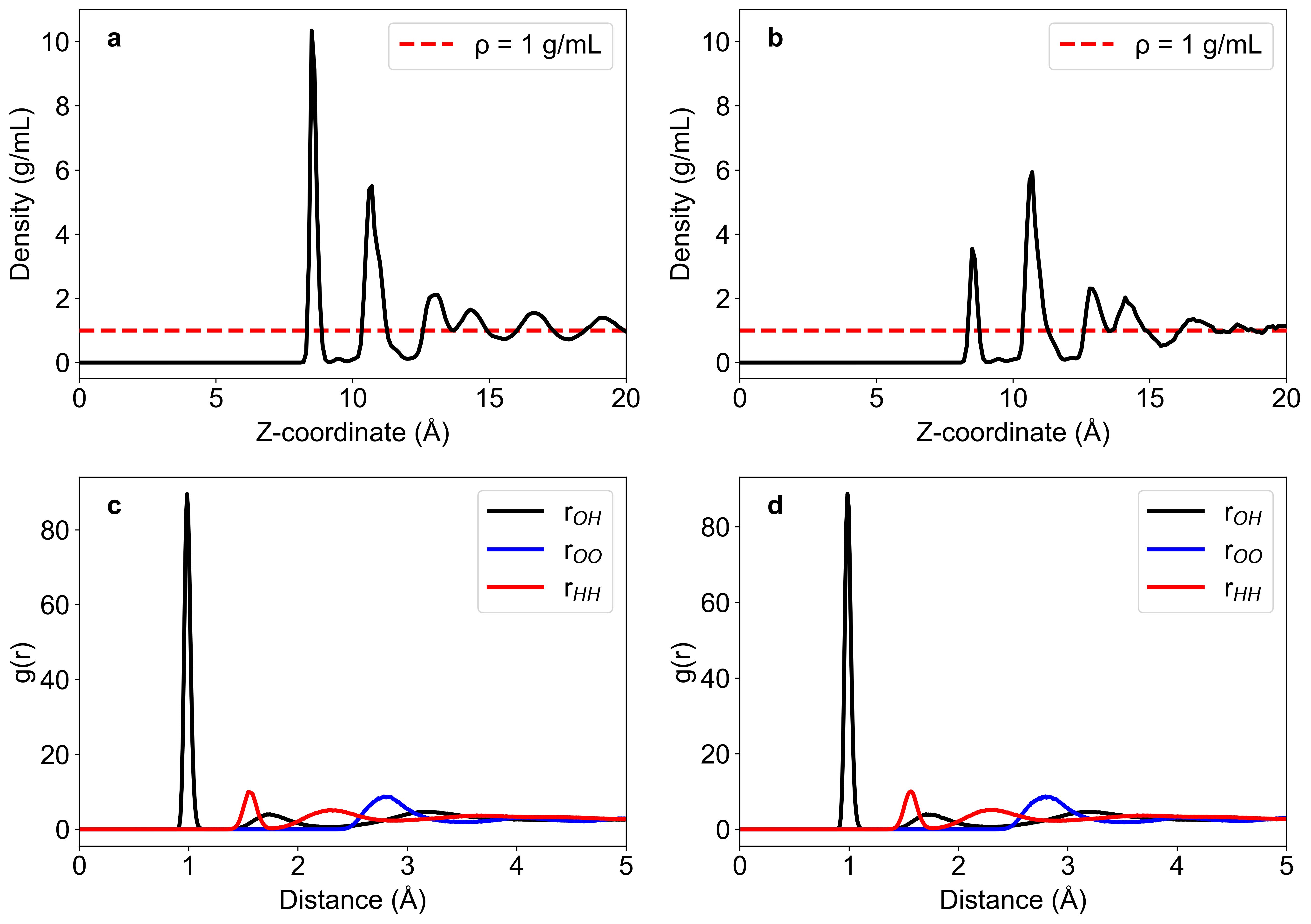}
\caption{(a-b) Density profiles and (c-d) RDFs of only water molecules were obtained from MLP-MD simulations of the (a,c) BTO(4x4)/128H$_{2}$O and (b,d) Ni@BTO(4x4)/128H$_{2}$O systems.}\label{figB2}
\end{figure}

\clearpage
\pagebreak

\section{MD and MetaD simulations}\label{secC1}

\begin{figure}[H]
\centering
\includegraphics[width=1\textwidth]{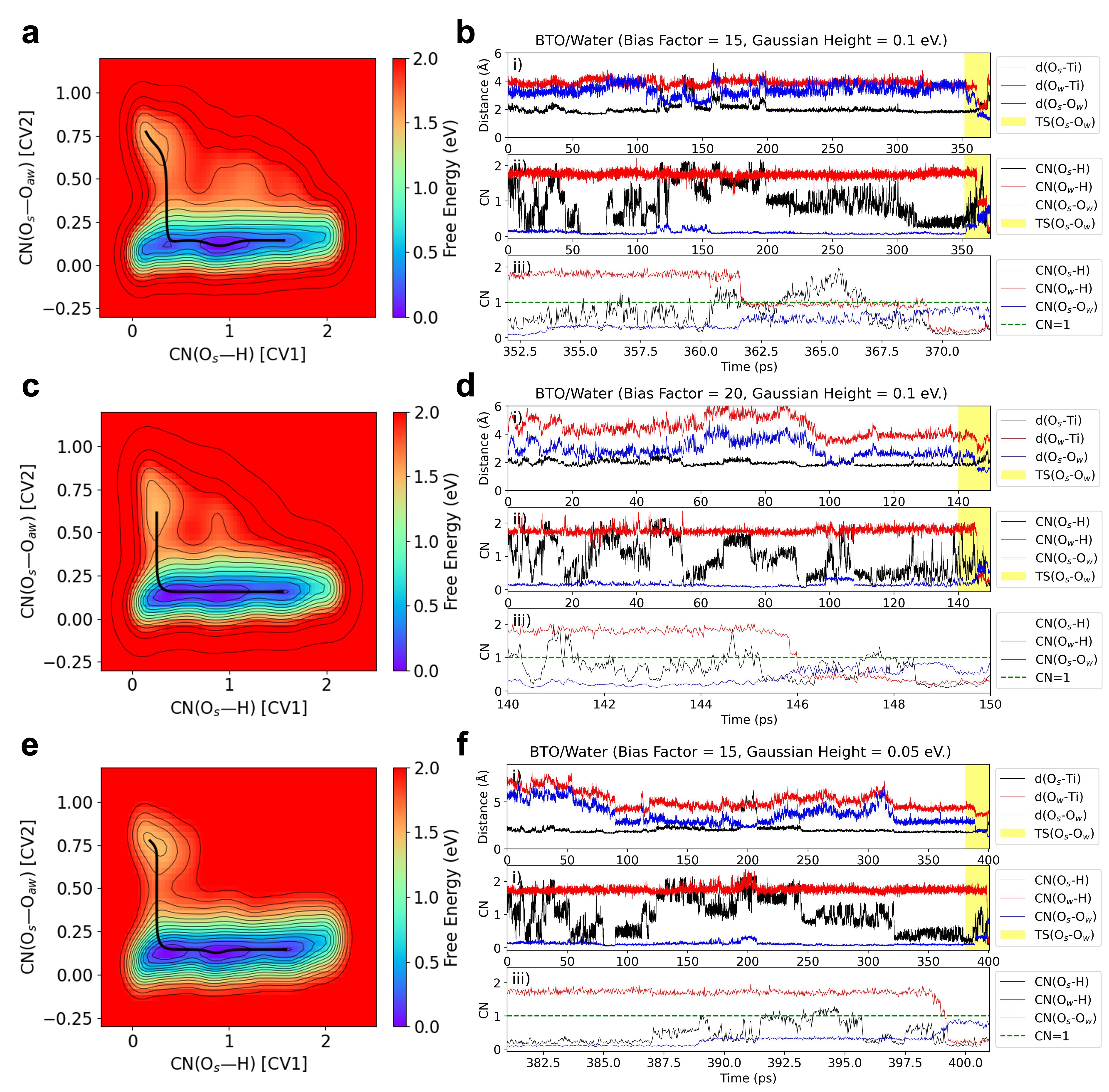}
\caption{Free energy surfaces and MetaD trajectory of the BTO/Water interface, using different bias factor and Gaussian height. (a-b) 15 and 0.1 eV. (c-d) 20 and 0.1 eV. (e-f) 15 and 0.05 eV. The MetaD trajectory includes the following analyses: (i) Distance measurements between specific atom pairs. (ii) Coordination numbers between selected atoms. (iii) Coordination numbers between selected atoms near the transition state, with the yellow-shaded area zoomed in for clarity. The free energy surfaces are calculated until O$_{2}$ formation to exclude the contribution of hydrogenated aqueous O$_{2}$ configurations that appear at longer simulation times.}\label{figC1}
\end{figure}

\begin{figure}[H]
\centering
\includegraphics[width=1\textwidth]{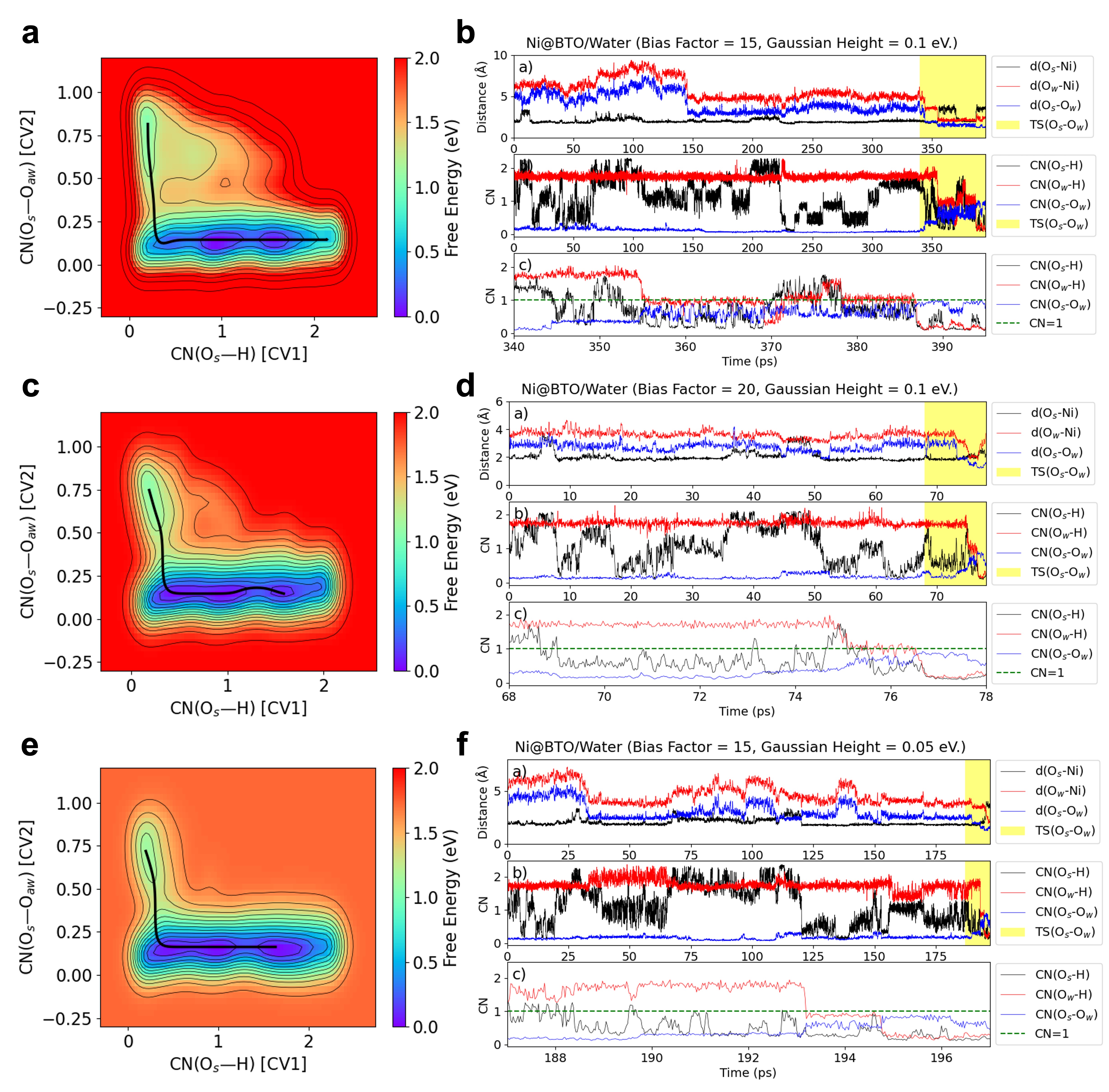}
\caption{Free energy surfaces and MetaD trajectory of the Ni@BTO/Water interface, using different bias factor and Gaussian height. (a-b) 15 and 0.1 eV. (c-d) 20 and 0.1 eV. (e-f) 15 and 0.05 eV. The MetaD trajectory includes the following analyses: (i) Distance measurements between specific atom pairs. (ii) Coordination numbers between selected atoms. (iii) Coordination numbers between selected atoms near the transition state, with the yellow-shaded area zoomed in for clarity. The free energy surfaces are calculated until O$_{2}$ formation to exclude the contribution of hydrogenated aqueous O$_{2}$ configurations that appear at longer simulation times.}\label{figC2}
\end{figure}

\begin{table}[h!]
\caption{OER kinetic barriers were derived from the free energy surfaces of the BTO/water and Ni@BTO/water systems. Independent MetaD simulations, with varied MetaD parameters and initial configurations, were performed to confirm the convergence of the kinetic barriers. The gaussian height and free energies is represented in unit of eV. The free energy surfaces are calculated until  O$_{2}$ formation. }\label{tabC1}%
\begin{tabular}{@{}lcccccccccc@{}}
\toprule
Model & Bias Factor & Height & $G_{H_{2}O}$ &$G^{\ddagger}_{H_{2}O\to OH}$ &$G_{OH}$ & $G^{\ddagger}_{OH\to O}$  & $ G_{O}$  & $G^{\ddagger}_{O\to O_{2}}$ &$ G_{O_{2}}$  \\
\midrule
\textbf{BTO}      &   &   &  &  &   &  \\
FES 1     & 15  & 0.10  & 0.00 & 0.03 & -0.28 & 0.01 & -0.10 & 1.43 & 1.21 \\
FES 2     & 20  &0.10  & 0.00 & 0.07 & -0.15 &0.11 &  -0.07 &  1.43 & 1.30 \\
FES 3     & 15  &0.05   & 0.00 & 0.08 & -0.13 & 0.04 & -0.11 & 1.58 & 1.34 \\
\midrule
Mean  &    &   & 0.00 & 0.06 & -0.19 &  0.05& -0.09 & 1.48 & 1.28\\
S.D.  &    &   & $\pm$0.00 & $\pm$0.02 & $\pm$0.07 & $\pm$0.04 & $\pm$0.02 & $\pm$0.07 & $\pm$0.05  \\
\midrule
\textbf{Ni@BTO}      &   &   &  &  &   &  \\
FES 1     & 15  & 0.10  & 0.00 & 0.19 & -0.04  & 0.26 & 0.20 & 1.31 & 0.99 \\
FES 2     & 20  &0.10   &  0.00 & 0.14 & -0.08 & 0.01 & -0.05 & 1.20 & 0.98 \\
FES 3     & 15  &0.05  &0.00	&	0.17&	-0.02& 0.07	&0.02 & 1.26 & 1.11\\
\midrule
Mean  &    &   & 0.00	& 0.17	&-0.05	&0.11	& 0.06 & 1.26&1.03\\
S.D.  &    &   & $\pm$0.00 & $\pm$0.02	& $\pm$0.02	& $\pm$0.10	& $\pm$0.10  & $\pm$ 0.00& $\pm$0.00\\
\botrule
\end{tabular}
\end{table}

\begin{figure}[H]
\centering
\includegraphics[width=0.5\textwidth]{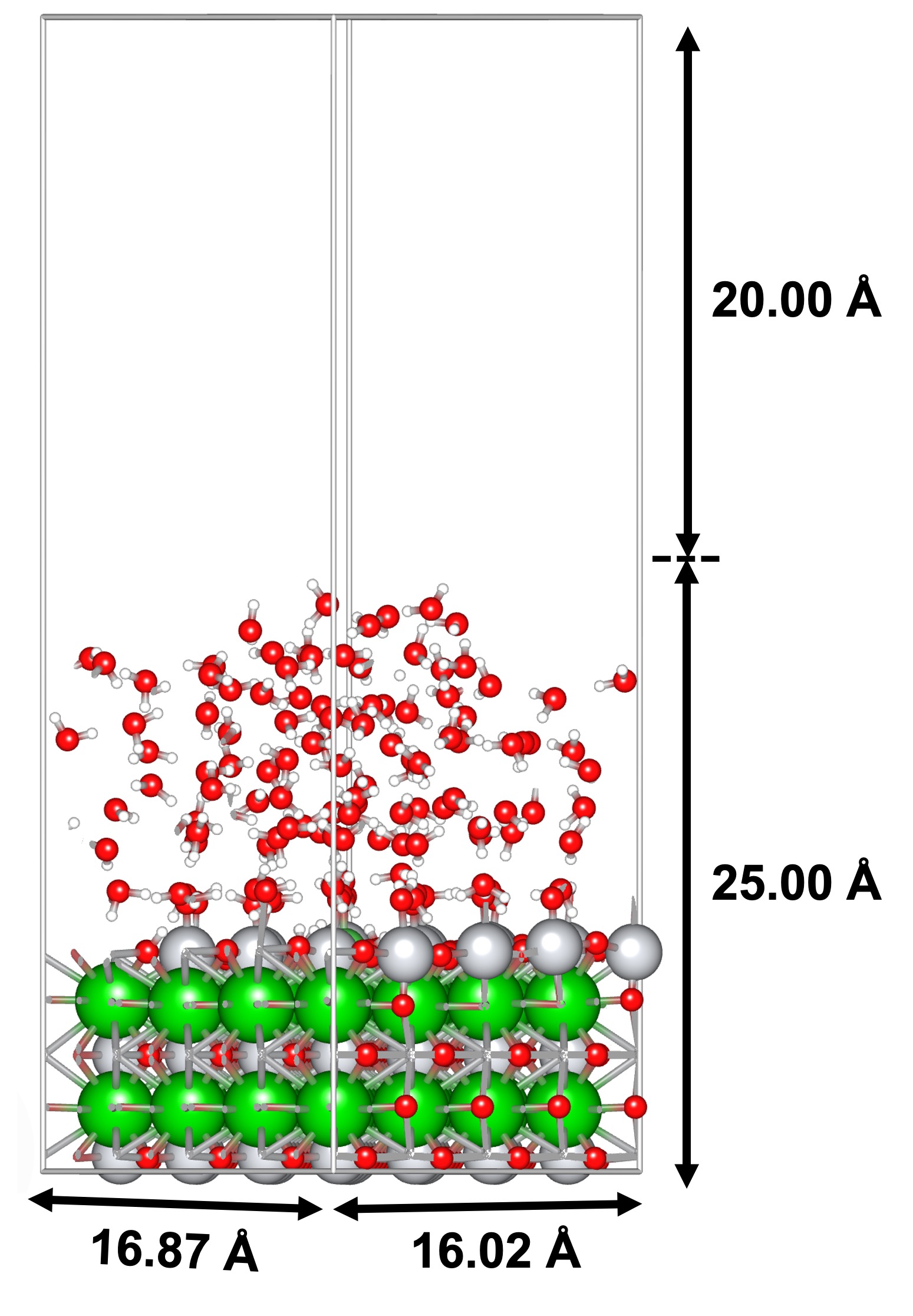}
\caption{The MD and MetaD simulation box of the slab model interface with water, consists of 208 atoms in the solid structure (4$\times$4 supercell) and 384 atoms in the solvent (128 water molecules).
}\label{figC3}
\end{figure}

\begin{figure}[!h]
\centering
\includegraphics[width=1\textwidth]{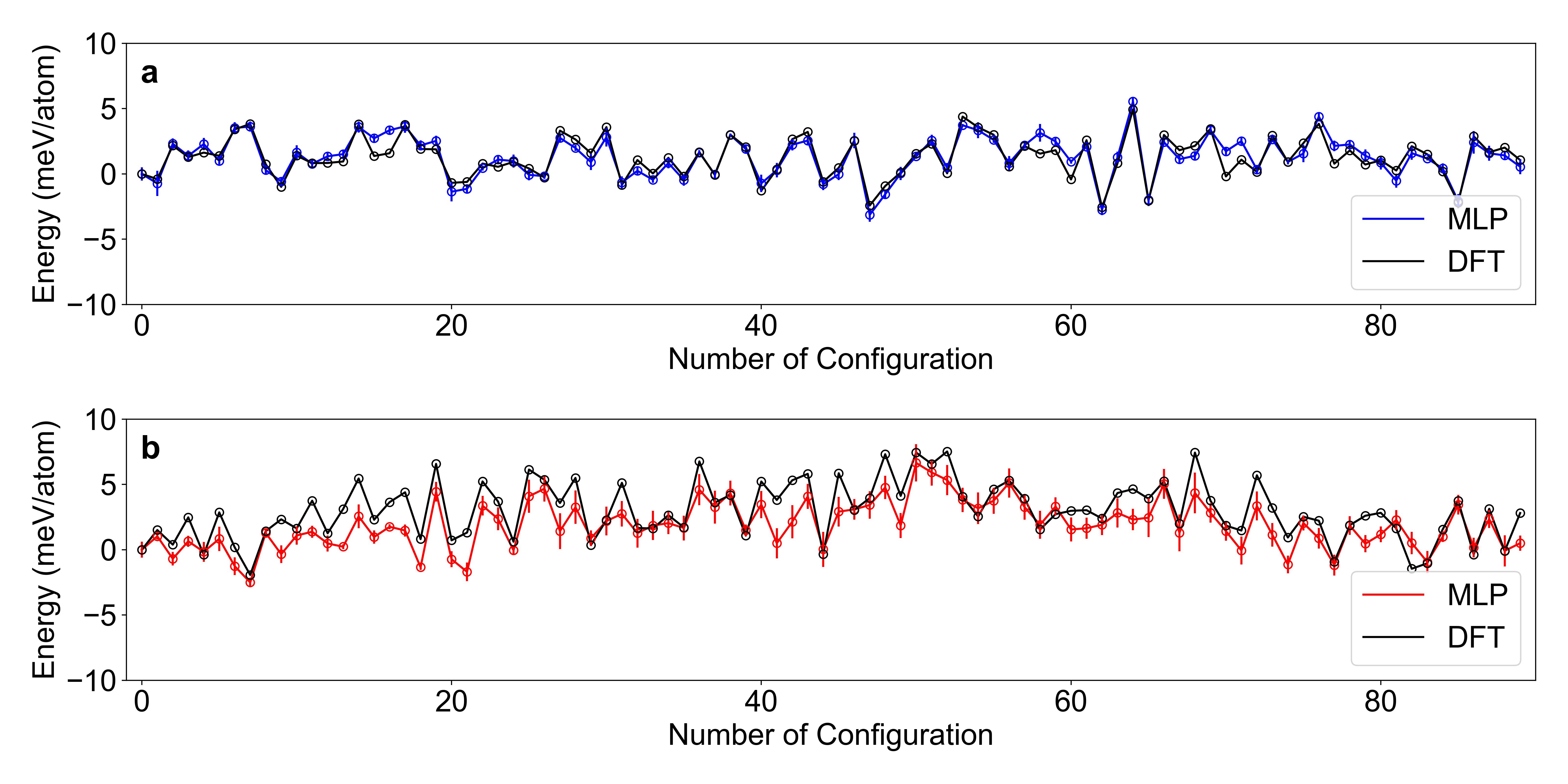}
\caption{A comparison of the relative total energies for structures sampled from MetaD trajectories (slab(4$\times$4)/128H$_{2}$O), used as an independent test set (a) BTO configurations (b) Ni@BTO configurations. The predicted energies were drawn from the  ensemble of MLPs in Table~\ref{tabB1}. The error bar indicates the standard deviation of the prediction. }\label{figC4}
\end{figure}

\pagebreak



\end{appendices}

\clearpage\pagebreak
\bibliography{sn-bibliography}

\end{document}